# Spin Logic Devices via Electric Field Controlled Magnetization Reversal by Spin-Orbit Torque

Meiyin Yang, Yongcheng Deng, Zhenhua Wu, Kaiming Cai, Kevin William Edmonds, Yucai Li, Yu Sheng, Sumei Wang, Yan Cui, Jun Luo, Yang Ji , Hou-Zhi Zheng and Kaiyou Wang

*Abstract*—We describe a spin logic device with controllable magnetization switching of perpendicularly magnetized ferromagnet / heavy metal structures on a ferroelectric (1-$x$)[Pb(Mg$_{1/3}$Nb$_{2/3}$)O$_3$]-$x$[PbTiO$_3$] (PMN-PT) substrate using current-induced spin-orbit torque. The devices were operated without an external magnetic field and controlled by voltages as low as 10 V applied across the PMN-PT substrate, which is much lower compared to previous reports (500 V). The deterministic switching with smaller voltage was realized from the virgin state of the PMN-PT. Ferroelectric simulation shows the unsaturated minor loop exhibits obvious asymmetries in the polarizations. Larger polarization can be induced from the initial ferroelectric state, while it is difficult for opposite polarization. The XNOR, AND, NAND and NOT logic functions were demonstrated by the deterministic magnetization switching from the interaction between the spin-orbit torque and electric field at the PMN-PT/Pt interface. The nonvolatile spin logic scheme in this work is simple, scalable, programmable, which are favorable in the logic-in-memory design with low energy consumption.

*Index Terms*—Electric field, logic, spin-orbit torque,

## I. INTRODUCTION

LOGIC and programmable memory devices are of great significance for information technology today. A spin logic function by spin-orbit torque (SOT) is a promising candidate to realize such devices with high integration and low power consumption. In heavy metal/ferromagnet systems, spin-orbit torques can be used to switch the perpendicularly magnetized ferromagnets [1]-[3]. The spin polarization due to SOT is directed perpendicular to the out-of-plane ferromagnetic easy axis, providing an additional degree of freedom to control the local magnetization switching [4] , [5] . This offers a way to construct logic functions in a single memory cell [6] and greatly favors the scaling of integration [7]. The SOT enables the separation of reading and writing channels in magnetic tunneling junctions, which will greatly increase the endurance of magnetic random access memory (MRAM) arrays. However, generally an in-plane magnetic field is required to realize the deterministic magnetization switching induced by SOT [8]-[11], and the direction of the in-plane field determines the chirality of the magnetization loop, *i.e.* whether a positive current favors "up" or "down" orientations of the magnetization. The requirement for an external field impedes high-density integration. Although lots of efforts have been made to realize magnetic field-free switching by SOT, including using wedged structures [12], [13], tilted anisotropy [14]-[16], exchange bias [17]-[20], interface engineering [21], stray fields [22], [23] or exchange coupling [24]-[27], the controllable chirality of the magnetization switching was lost for most reports. All-electrical manipulation of magnetization while maintaining the chirality control, which is crucial for programmable storage, still requires work.

Recently, electric field control of deterministic current-induced magnetization switching was demonstrated in a hybrid ferromagnetic-ferroelectric structure [28]. By polarizing the PMN-PT substrate with an applied voltage, deterministic switching was achieved by electrical current in the absence of an external magnetic field. The chirality of the magnetic switching loop can also be controlled by changing the sign of the voltage applied to the PMN-PT substrate. However, the 500 V required to polarize the PMN-PT substrate is too large for practical applications. Also, logic functionality was not demonstrated in Ref. [28] due to the requirement of high voltage. Here, we describe devices in which the distance between the electrodes on the PMN-PT is reduced from 1 mm to 100 $\mu$m, so that the voltage required is greatly reduced. By applying voltage pulses to the PMN-PT and current pulses to the heavy metal/ferromagnet device, we demonstrate multiple logic functions using a single magnetic unit.

## II. EXPERIMENTS

Pt (3 nm)/Co (0.9 nm)/AlO$_x$ (2 nm) films were deposited on top of (001)-oriented PMN-PT ferroelectric substrates with thickness of 0.5 mm by magnetron sputtering at room temperature. Pt, Co, and Ni were sputtered at a pressure of 0.8 mTorr with deposition rates of 1.3 nm/min, 0.7 nm/min, and

This work was supported by National Key R&D Program of China No. 2017YFB0405700 and 2017YFA0303400. This work was supported also by the NSFC Grant No. 11604325, 11474272 and 61774144. The project was sponsored by Chinese Academy of Sciences, grant No. QYZDY-SSW-JSC020, XDA18000000, XDPB12 and XDB28000000.

M. Yang, Y. Deng, Y. Li, K. Cai, Y. Sheng, Y. Ji, Z. Zhi and K. Wang are with SKLSM, Institute of Semiconductors, CAS, P. O. Box 912, Beijing 100083, China (e-mail: kywang@ semi.ac.cn).
M. Yang, Z. Wu, S. Wang, Y.Cui and J. Luo are with Key Laboratory of Microelectronic Devices and Integrated Technology, Institute of Microelectronics, Chinese Academy of Sciences, Beijing 100029, China.
K.W.Edmonds is with School of Physics and Astronomy, University of Nottingham, Nottingham NG7 2RD, United Kingdom.
K. Wang is with Center of Materials Science & Optoelectronics Engineering , Center for Excellence in Topological Quantum Computation, University of Chinese Academy of Sciences, Beijing 100049 , P. R. China, and Beijing Academy of Quantum Information Sciences, Beijing 100193, China.

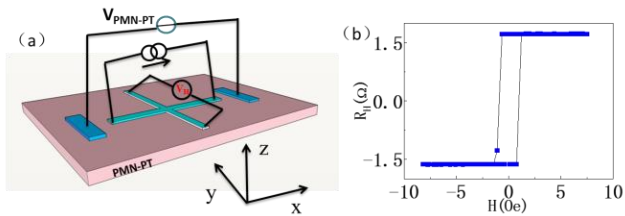

Fig. 1. (a) The schematic device structure for Pt/Co/AlOx sample. The current and voltage channels of the Pt/Co/AlOx film are 2 um wide (light blue), while the voltage contacts to the PMN-PT substrate are separated by 100 um (dark blue) (b) The anomalous Hall resistance loop of the device under a perpendicular magnetic field.

2.52 nm/min respectively. The Pt/Co/AlO$_x$ film was subsequently patterned into an array of Hall bar devices by electron beam lithography (EBL) and etching techniques. A Keithley 2602 current source and a Keithley 2182A nanovoltmeter were used for the d.c. Hall voltage measurements. The ferroelectric simulation was done by the technology computer aided design (TCAD) software using the package of global TCAD solutions (GTS).

### III. RESULTS AND DISCUSSIONS

#### A. Basic magnetic properties

The device structure of the Pt/Co/AlO$_x$ film is shown in Fig. 1(a). The channels for current injection and Hall voltage detection are 2 $\mu$m wide. The separation of the voltage electrodes on the PMN-PT substrate is 100 $\mu$m, which is one-tenth of the distance in the previous work [28]. The hysteresis loop of the anomalous Hall resistance ($R_H$) with the magnetic field applied in the out-of-plane direction is presented in Fig. 1(b). The square hysteresis loop shows that the Pt/Co/AlO$_x$ sample has perpendicular anisotropy with a switching field of 2 Oe.

#### B. The Logic functions

The electrical current switching behavior of the Pt/Co/AlO$_x$ sample after applying a voltage to the PMN-PT substrate is shown in Fig. 2. The minimum voltage to observe the deterministic switching without a magnetic field is 10 V, which was applied first on the PMN-PT substrate, and then removed before the electrical measurements. Figure 2(a) presents the $R_H$-I loop after applying the voltage of 10 V. Consistent with our previous work [28], the deterministic switching was achieved but with much less voltages. The decrease of the voltage is an important step towards the application of SOT-induced programmable memory [29].

For –10 V applied to the PMN-PT, no deterministic switching was observed, indicating no preferred polarization under the device. The reversed deterministic switching was observed after applying –30 V (Fig. 2(b)). The size of the Hall resistance signal indicates that nearly 80% of the magnetic domains exhibits deterministic switching. Fully deterministic switching did not show up when further raising the voltage.

The switching behavior of the Pt/Co/AlO$_x$ device is closely related to the sequence of the applied voltage on the PMN-PT substrate. We found that the deterministic switching could be achieved under a small voltage from the virgin state of the PMN-PT. Then it requires larger voltage to reverse the deterministic switching loop, and only partial reversal is achieved. To explain this phenomenon, we conducted the ferroelectric hysteresis loop simulation as is shown in Fig.1 (c). The saturated major loop is very symmetric, while the initial minor loop was obviously asymmetric. The initial remanence polarization (point (1)) is larger than that of point (4) after applied with the same opposite electric field. Thus, due to the surface local electrodes on the bulk PMN-PT substrate, the devices are operated at the minor zone so that the magnetization loop exhibits asymmetric between 10 V and -30 V voltages on the PMN-PT substrate. The asymmetric behaviors could be improved by using patterned ferroelectric thin film with low coercivity field in future.

A XNOR logic function was demonstrated as illustrated in XNOR gate, which is realized using only one memory unit.

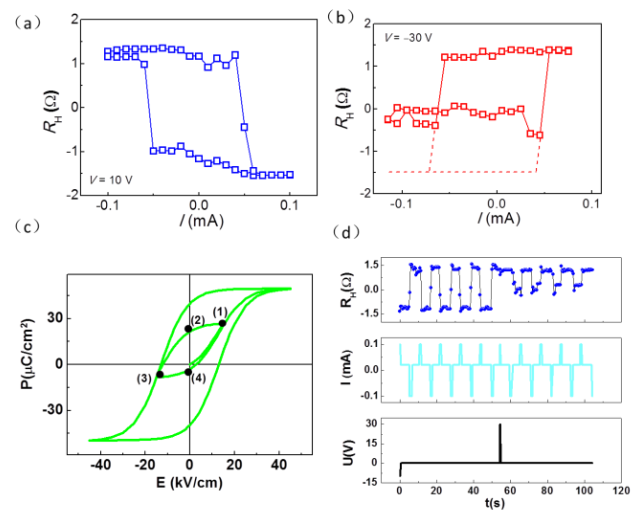

Fig. 2. (a,b) Current-induced magnetization switching after a voltage was applied across the PMN-PT substrate for the Pt/Co/AlO$_x$ sample, (a) 10 V, (b) -30 V. (c) The simulated ferroelectric Polarization vs. electric field hysteresis loop. (d) Demonstration of the XNOR logic function, showing the Hall signal $R_H$ after current pulses I applied to the Pt/Co/AlO$_x$ device and voltage pulses U applied to the PMN-PT.

Fig. 2(d). The voltage applied to the current channel and the PMN-PT electrodes are defined as the two inputs, and the Hall resistance signals are the outputs. Positive and zero Hall resistance are defined as "1" and "0", respectively. Current pulses of ±0.1 mA with the duration of 0.5 s were injected and then 0.02 mA constant current was applied to read the Hall resistance. Voltages of 10 V and –30 V were applied to the PMN-PT electrodes and removed before the current pulses were injected. First, after the 10 V pulse (input "1"), the magnetization followed the switching loop in Fig. 2(a), where the positive current (input "1") resulted in the negative Hall resistance (output "0") and negative current led to positive Hall resistance (output "1"). After the –30 V pulse (input "0"), the magnetization switching was reversed, so that the positive current (input "1") and negative current (input "0") generated positive resistance (output "1") and nearly zero resistance (output "0"). The switching behavior acts as the function of

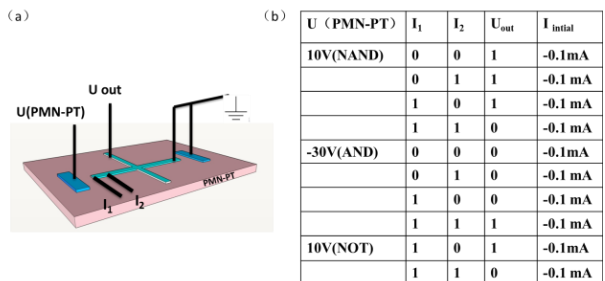

Fig. 3(a) The schematic device structure for Logic gates, (b) the truth table of Logic gates

Besides the XNOR logic function, AND, NAND and NOT logic gates are proposed. The logic gates setup and true table were shown in Fig. 3. The two inputs are the currents that applied to the Hall channel. The "0" represents 0 mA and the "1" represents "0.03 mA". The magnetization are initialized before every operation by injecting a large current (e.g. -0.1 mA). NAND logic function gate can be achieved when applying 10 V to the electrodes on PMN-PT. Only on one condition that the two inputs are "1" which exceeds the critical switching current can reverse the magnetization, resulting a lower voltage output ( $U_{out}$=0), which is a NAND logic function. Besides, if we fixed $I_1$ to "1" every time (which can be regarded as the operating current) and varied the input $I_2$, the outputs are always opposite to $I_2$, resulting a NOT gate functions. If we applied -30 V to the electrodes on PMN-PT, only two "1" inputs can set the output to "1", which is a AND logic. Thus, we can achieve three different logic functions using the same structure.

### C. The features of the new logic scheme

The proposed AND, NAND and NOT logic gates are concatenable since the input and output bits are encoded with the same physical quantities: current amplitude. For Hall devices, the output current are realized by anomalous Hall voltage, which transferred into current to the next logic gates. The gain of the logic by Hall bar is small but can be improved using a magnetic tunnel junction. The controllable deterministic switching methods in this paper is compatible to

TABLE I
COMPARISON OF LOGIC DEVICES BY DIFFERENT METHODS

|  | This work | Ref. [30] | Ref.[31] | Ref.[7] |
|---|---|---|---|---|
| E | ~2fJ | ~2fJ | ~1fJ | ~1.5 pJ |
| Non-volatile | yes | no | no | yes |
| Transistors | 3 | 6~12 | 1 | 3 |
| C/M | M | C | M | M |

E represents the energy consumption and C/M symbolizes the CMOS or ferromagnetic logic gate.

MTJ, in which case the output can be directly as the input to the other devices without amplifier. The logic output are almost linear before the switching of the magnets, since the magneto-resistance hardly changes with the applied voltage or current. Since the logic gate only use one memory cell and the magnets are stable below the Curie temperature, the necessity of feedback elimination is greatly reduced.

Table I summarized some averaged features of logics functions by CMOS, other spintronics logics and our work. The energy consumption of this work was calculated with the speed of 5 ns using 500 nm Hall bar. Due to the low switching current, the energy consumption of our logics is lower than previous spintronic devices and then comparable with the CMOS logic gates. Further more, this device is also a memory, so that no static energy are consumed. Also, the different logic gates(AND, NAND, XNOR and NOT) can be achieved in the same cell with less transistors needed, which is favorable in the future logic-in-memory devices.